\documentclass[reprint,
 amsmath,amssymb,
 aps,graphicx
]{revtex4-2}

\usepackage{graphicx}
\usepackage{dcolumn}
\usepackage{bm}
\usepackage[dvipsnames]{xcolor}
\DeclareMathOperator*{\E}{\mbox{E}}
\DeclareMathOperator*{\V}{\mbox{Var}}

\begin{document}

\preprint{APS/123-QED}

\title{Quantified Advantage of Ghost Imaging over Regular Imaging}

\author{Anjaneshwar Ganesan$^{1}$}
\author{Herman Batelaan$^{2}$}
\affiliation{$^{1}$Georgia Institute of Technology School of Physics, Howey Physics Bldg, 837 State St NW, Atlanta, GA 30332, anjan.ganesan@gatech.edu  \\ $^{2}$Department of Physics and Astronomy, University of Nebraska, Jorgensen Hall, Theodore, 855 N 16th St Room 208, Lincoln, NE 68588, hbatelaan2@unl.edu}




\begin{abstract}
{\bf Abstract.} Ghost imaging is a remarkable technique where light that never interacts with an object is detected with a camera and still the image of the object is recorded. The method relies on the use of correlated light and an additional ``bucket" detector. Ghost imaging has been used in archaeology, bio-medicine, for seeing through turbid media, and promises X-ray imaging improvements, amongst many other applications. However, the advantage of ghost imaging over regular imaging can be difficult to quantify. For classical ghost imaging of a single pixel aperture (the object), we find a closed analytic expression for the signal-to-noise ratio using basic statistics. We find that this signal-to-noise ratio can exceed that of regular imaging with the same exposure of the aperture when the detectors are sufficiently noisy, illustrating a simple and quantifiable advantage. Numerical simulation confirms the theoretical analysis.

\end{abstract}

\maketitle

\section{Introduction}
Ghost imaging (GI) offers the promise to see through “cloudy” objects \cite{yuan2022unsighted} and to lower the exposure of the imaged object \cite{batelaan2022shining}. The latter is especially important for soft biological tissue that can be damaged under exposure. An inconvenience is that the method is technically more involved, and it is hard to find simple examples where ghost imaging can quantitatively be shown to be advantageous, in comparison to regular imaging (RI). To that end, we consider the simplest example we can find. Thus, in this paper, we study a single pixel hole object and judge its image quality by the Signal-to-Noise Ratio (SNR). For a fair comparison, the exposure of the object is kept the same for ghost and regular imaging.
For a clear accessible introduction on how to perform ghost imaging we refer the reader to, for example, Ref. \onlinecite{basano2007ghost}.

Comprehension of the basic idea of how classical ghost imaging works, including computational ghost imaging, is described in many textbooks. In addition, there are many fascinating examples of the demonstration of ghost imaging. It is harder to answer the question where the advantage of ghost imaging comes from. Why not use a camera and take an image instead of using a technically more involved method? We believe that the answer is not straightforward and has multiple facets. Many experiments study the advantages of ghost imaging. For example, imaging through turbid media can give better contrast than regular imaging\cite{PhysRevLett.110.083901}. Exposure claims made for ghost imaging with X-rays \cite{zhang2018tabletop} have been debated \cite{PhysRevA.103.033503} illustrating the challenge of quantifying an advantage.
A comparison of the SNR of ghost imaging with light sources of a different nature (thermal, low flux, and high flux entangled) was made \cite{PhysRevA.79.023833}. In very specific situations (for example with sparse representation), X-ray and electron ghost imaging using multiplexing offers SNR improvements,\cite{lane2020advantages} though the exposure of the object is increased. Even if this is not the topic of our present discussion, this type of ghost imaging can still be advantageous for using more of the incident flux of expensive-to-operate sources.  Multiplexing is not advantageous when using photon counting detection. Foundational experiments, like the double-slit experiment, can also benefit from ghost imaging \cite{aspden2016video}. The above studies share that no quantitative comparisons to regular imaging were made, and that was also not their intention.

Closer to our approach is a recent study,\cite{li2020performance} where the SNR of regular imaging and computational ghost imaging was compared by varying the exposure of the object and by varying the object's transmission area. Our ghost imaging analysis is different in that it considers camera noise and quantum efficiency, while the analytical results can be used to identify the regimes where ghost imaging is advantageous. Ghost imaging makes it possible to obtain high contrast images when using an average of fewer than one detected photon per image pixel,\cite{morris2015imaging} which is relevant in view of the result we obtain that low photon numbers provide the greatest advantage. 

Because the presence or absence of a photon is numerically described by a ``1" or a ``0", our analysis is also applicable to the presence or absence of a light pulse. Correspondingly, the meaning of camera noise can be photon dark count or noisy room light. We believe that this may be of some use, especially given that it was shown that the cost for experiments to do ghost imaging can be lowered,\cite{aguilar2019low} and a new single pixel camera for ghost imaging was developed \cite{kuusela2019single}.   

We calculated the SNR for 1-D ghost and regular imaging. The object consists of an aperture with a size of one pixel. Our statistical analysis provides a closed expression. It is well known that the quality of ghost imaging depends on the object to be imaged\cite{gatti2004correlated,shapiro2012physics,moreau2018resolution}. However, ``quality" is hard to quantify. Thus even if limited in scope, the object of study chosen is a single pixel hole. The main benefit of studying the simplest possible object is to show a clear quantifiable ghost imaging advantage and our closed analytic expression provides insight into the scaling of that advantage. In addition, our approach provides a means for a complete ghost image to quantify what part of advantage is a single pixel one. We show that when the detector noise is increased to reduce the SNR of regular imaging to below one, ghost imaging is protected to yield an SNR that is better by an order of magnitude. This result, which is consistent with that of Ref. \onlinecite{morris2015imaging}, we confirm by simulation.

For ghost imaging, spontaneous parametric down-conversion (SPDC) entangled photon sources can be used \cite{lin2021efficient}. The use of SPDC sources is convenient and common for correlated photon ghost imaging \cite{aspden2013epr}. In the supplementary materials, we apply the analytic expressions for photon numbers of typical correlated photon sources. For quantum efficiencies of photon detectors above 0.4, we find a significant improvement of the SNR. That the advantage works for each pixel of a ghost image for low photon numbers motivates application when low exposure is required, such as in the area of imaging soft bio-tissue. 
The development of environmental scanning electron microscopy \cite{danilatos1988foundations} and thin fluid cells for transmission electron microscopy \cite{ross2015opportunities} makes imaging of biological tissue in their natural environment possible. Even so, electron exposure can kill living cells during the imaging process \cite{de2016live}. Similarly, photon exposure can affect living processes \cite{cole2014live} and even two photon exposure can benefit from reducing exposure \cite{landes2021quantifying}. Recent debates \cite{wojcik2015graphene, kennedy2017gene, de2016live, koo2020live} underline that capturing living cells in action is a holy grail in electron imaging, and strongly motivates all techniques that reduce the exposure of the object.

\section{Historical Background and Context}
A schematic of ghost imaging \cite{pittman1995optical} is shown in figure 1. A source emits two photons with equal and opposite momenta (see Fig. 1 inset (a)). When a photon is detected passing through the slit, we would expect to observe a photon at a correlated camera position, as the photon momenta are correlated. One records the image at the camera to confirm this. (This is a repetitive coincidence detection of both photons). Y.H. Kim et \textit{al.} performed this ``bi-photon" ghost imaging\cite{kim1999experimental}. Further expanding on this idea, if the slit is narrowed, then this can be observed by the camera not looking at the slit (see Fig. 1 inset (b)). This captures the basic idea of ghost imaging: observe an object (the slit) while not looking at it, at least not with the camera. A thorough discussion of ghost imaging is provided in chapter 16.4 of Ref. \onlinecite{al2016optics}.  Ghost imaging is a comparatively new imaging technique and more complex then regular imaging. It is thus natural to compare this technique to regular imaging and ask whether there is a regime where ghost imaging is advantageous over regular imaging.
\begin{figure}[h]
   \includegraphics[width = \linewidth]{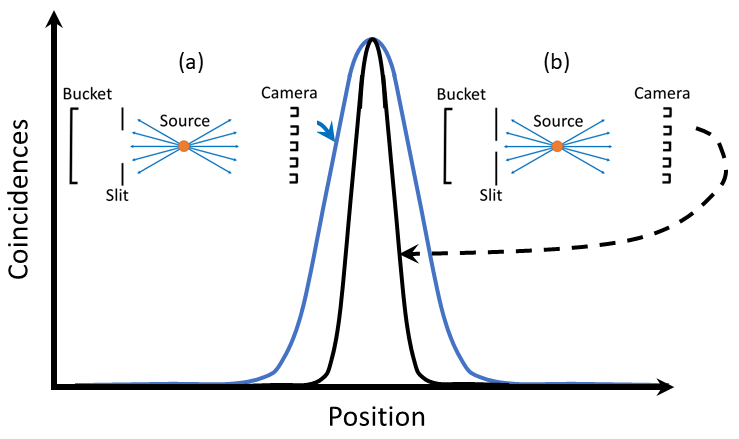} 
   \caption{A photon pair is released from a source, so that their momenta are correlated. Or alternatively, two correlated light beams are emitted back-to-back. The left inset (a) shows the image after repeated coincident detections of the bucket and camera. The observed peak is the ghost image of the slit. Right inset (b) shows the same except the slit is narrowed. The camera does not detect any light that passed through the slit illuminating the ``ghostly" nature of the image.} 
  \label{fig:setup}
\end{figure}

Pittman et \textit{al.} show ghost imaging for the first time, using the spatial correlations produced by parametric down-conversion \cite{pittman1995optical}. The effect was thought to be associated with the entanglement of the photons, but it was shown later that it is possible to perform ghost imaging with classically correlated light beams \cite{bennink2002two}. In other words, the above described versions of ghost imaging are called classical ghost imaging. In addition, it is not necessary to use a camera at all, when the spatial pattern of the emitted light is controlled and known at the source. This implies that the image at the camera is known without measurement, and by using that image, the same ghost image is obtained. This technique is called computational ghost imaging. Ghost imaging has spawned further progress in quantum imaging, notably the recent work by Lemos et \textit{al.} \cite{lemos2014quantum}  Interference between the photons from two down-conversion crystals, pumped with the same source, allows one to retrieve the image even if the beam that interacts with the object is not detected, achieving a feat that is perhaps even more ghostly than ghost imaging.

\section{Theory}
We do an SNR analysis of a regular imaging system and a classical ghost imaging system to quantify the quality of the image, while making sure that the exposure of the object is the same for both.  We start with the derivation of the signal for ghost imaging. Consider the setup in figure 2. A spontaneous down-conversion crystal (SPDC) generates two photons. One of these two goes to the camera and the other to the bucket detector such that they are spatially correlated. The number of produced photons received by the camera is given by $I_n(x_i)$ at the ``pixel" position $x_i$, where $i$ runs from $1$ to the number of pixels $P$. One such image is the $n$-th ``realization." The number of photons in each pixel in the spatial distribution is assumed to be given by a Poisson distribution $X_{i,n}(\lambda)$, where $\lambda=Rt_{w}$ (R is the rate of detected photons per pixel, and $t_{w}$ is the time window). The object is placed in front of the bucket detector. The transmission of the object per pixel at position $x_i$ is given by  $T(x_i)$. For each realization $n$, the bucket signal $B_n$ is obtained by summing over the pixel position, while an individual ghost image is $I_n(x_i)B_n$. Summing over $N$ realizations leads to the final ghost image $G(x_i)$:  
\begin{align}
    &I_n(x_i)=X_{i,n}(\lambda)\\
    &B_n=\sum_{i}I_n(x_i)T(x_i)\\
    &G(x_i)=\sum_{n=1}^{N}B_nI_n(x_i).
\end{align}
 
For regular imaging, we replace the bucket detector with a camera and take the sum of $I_n(x_i)T(x_i)$ over $N$ realizations:

\begin{align}
    &R(x_i)=\sum_{n=1}^{N}T(x_i)I_n(x_i).
\end{align}

In our case, regular imaging is shadow imaging, such as x-ray imaging used in radiology. 


\begin{figure}[ht]
   \includegraphics[width = \linewidth]{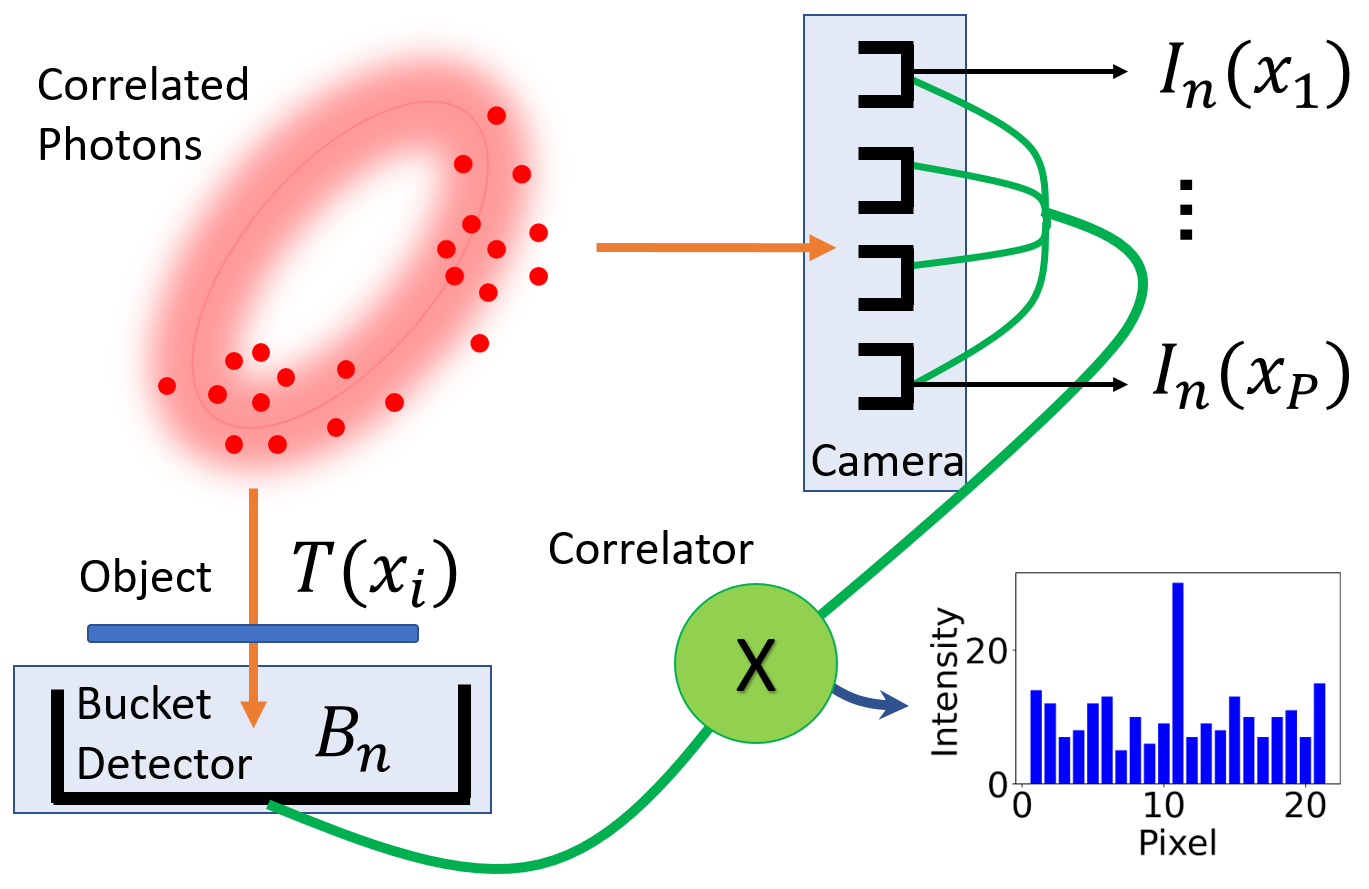} 
   \caption{Schematic Setup. The correlated source sends spatially correlated photons to the object and the camera's pixels. After the photons expose the object, the bucket detector collects the transmitted photons, and the correlator multiplies the bucket detector signal with the pixelated image from the camera. This gives us a ghost image for the $n$-th realization. An example ghost image summed over many realizations is shown in the blue bar graph. If we remove the camera and correlator and replace the bucket detector with the camera, we get the setup for regular imaging. If we switch the roles of the bucket detector and camera, we get gated imaging.} 
  \label{fig:GIsetup}
\end{figure}

\subsection{Simulation}

The object we consider is a single pixel hole, i.e., a slit where the size of the slit is exactly one pixel wide (all pixels in the object are black except for one fully transparent white pixel). The image of a single pixel hole object consists of one data point in the location $x_{i=m}$ separated from a background of data points in the locations $x_{i\neq m}$. To get the signal, we take the average intensity of the background over the number of  background pixels and subtract it from the intensity at point $x_{i=m}$. Thus, the signals of both regular and ghost imaging needed to compute the signal-to-noise ratio is denoted by $S_R$ and $S_G$ respectively, are given by
\begin{align}
    &S_R = R(x_m)-\sum_{i=1}^{P-1}\frac{R(x_{i\neq m})}{P-1}\\
    &S_G = G(x_m)-\sum_{i=1}^{P-1}\frac{G(x_{i\neq m})}{P-1},
\end{align}
where $P$ is the total number of pixels in the camera. 

To get the noise in the signals $S_R$ or $S_G$, one measurement of the signal is given by Eqs. 5 and 6. The measurement is repeated $K$ times $(k=1,...,K)$ and $\overline S$ is the average signal over $K$ measurements. Then, the noise in the signal, given by $\sigma_S$, is given by

\begin{align}
\sigma_S = \sqrt{\frac{1}{K} \sum_{k=1}^K (S_k - \overline{S})^2}.
\end{align}

Evaluating Eqs. 1 to 7 will give us the SNR for ghost and regular imaging. By using a Python simulation, we investigate how the SNR for both varies with $\lambda$ in Eq. 1. We are interested in varying $\lambda$ because that determines how much we expose the object $T(x_i)$. Since we get the image over $N$ realizations, and $\lambda$ is the average exposure of the object per realization, we get the same exposure $N\lambda$ for both regular and ghost imaging. 

\subsection{Analysis}

To verify the simulated result, we derive the SNR for regular and ghost imaging. The single pixel hole object is described by 
\begin{equation}
  T(x_i)=\left\{\begin{array}{ll}
    1, & i=m;\\[5pt]
    0, & i\neq m.
  \end{array}\right.
\end{equation}
The source intensity distribution that is sent to the bucket detector is identical to the one sent to the camera. This distribution is given in terms of photons per pixel by
\begin{equation}\label{}
    I(x_i)=X_i(\lambda_s).  
\end{equation}
$X_i(\lambda_s)$ is a random Poisson number dependent on $\lambda_s$ characterising the source. It is not sufficient to lower $\lambda_{s}$ below one to obtain a ghost imaging advantage over regular imaging (see Results section). After introducing detector dark counts, this situation changes. Any detector, even with the absence of light or particles, may display random counts due to, for example, cosmic rays or the electronic shot noise. Since we see these counts in the dark, they are called detector dark counts. The dark count in the detectors per pixel is given by  
\begin{equation}\label{}
    J(x_i)=Y_i(\lambda_d).  
\end{equation}
$Y_i(\lambda_d)$ is a random Poisson number dependent on $\lambda_d$.
The bucket detector signal is
\begin{align}
    \sum_{i}T(x_i)I(x_i)+J(x_b)&=I(x_m)+J(x_b)\nonumber \\
    &=X_m(\lambda_s)+Y_b(\lambda_d),\label{}
\end{align}
where $J(x_b)=Y_b(\lambda_d)$ is the bucket detector dark count. The ghost image over $N$ realizations is given by 
\begin{align}
    G(x_{i\neq m}) &= \sum_{n=1}^{N}(I(x_{m,n})+J(x_{b,n}))(I(x_{i\neq m,n})\nonumber\\
    &+J(x_{i\neq m,n}))\nonumber\\
    &=\sum_{n=1}^{N}(X_{m,n}(\lambda_s)+Y_{b,n}(\lambda_d))(X_{i\neq m,n}(\lambda_s)\nonumber\\
    &+Y_{i\neq m,n}(\lambda_d))\\
    G(x_{i=m}) &= \sum_{n=1}^{N}(I(x_{m,n})+J(x_{b,n}))(I(x_{m,n})+J(x_{m,n}))\nonumber\\
    &= \sum_{n=1}^{N}(X_{m,n}(\lambda_s)+Y_{b,n}(\lambda_d))(X_{m,n}(\lambda_s)\nonumber\\
    &+Y_{m,n}(\lambda_d)),
\end{align}
If the bucket detector were to be replaced by a camera, the regular image is given by
\begin{align}
    R(x_{i\neq m}) &= \sum_{n=1}^{N}[T(x_{i\neq m})I(x_{i\neq m,n})+J(x_{i\neq m,n})]\nonumber\\
    &=\sum_{n=1}^{N}Y_{i\neq m,n}(\lambda_d)\\
    R(x_{i=m}) &= \sum_{n=1}^{N}[T(x_{i=m})I(x_{m,n})+J(x_{m,n})]\nonumber\\
    &= \sum_{n=1}^{N}[X_{m,n}(\lambda_s)+Y_{m,n}(\lambda_d)].
\end{align}
By using Eqs. 5, 14, and 15, the signal of regular imaging measured with $P$ pixels over $N$ realizations is given by
\begin{align}
    S_R =& \sum_{n=1}^{N}\left[X_{m,n}(\lambda_s)+Y_{m,n}(\lambda_d)\right]\nonumber\\
    & -\sum_{n=1}^{N}\left[\sum_{i=1}^{P-1}\frac{Y_{i\neq m,n}(\lambda_d)}{P-1}\right].
\end{align}
By using Eqs. 6, 12, and 13, the signal of the ghost image measured with $P$ pixels over $N$ realizations is given by
\begin{align}
    &S_G = \sum_{n=1}^{N}\left[(X_{m,n}(\lambda_s)+Y_{b,n}(\lambda_d))(X_{m,n}(\lambda_s)+Y_{m,n}(\lambda_d))\nonumber\right]\\
    &-\sum_{n=1}^{N}\left[\sum_{i=1}^{P-1}\frac{(X_{m,n}(\lambda_s)+Y_{m,n}(\lambda_d))X_{i\neq m,n}(\lambda_s)}{P-1}\right]\nonumber\\
    &-\sum_{n=1}^{N}\left[\sum_{i=1}^{P-1}\frac{(X_{m,n}(\lambda_s)+Y_{m,n}(\lambda_d))Y_{i\neq m,n}(\lambda_d)}{P-1}\right].
\end{align}   

Note that the parametrization, in particular $\lambda_s$, allows the exposure for regular imaging to be chosen identical to that for ghost imaging. 
The SNRs of regular imaging and ghost imaging are given by 
\begin{align}
    \text{SNR} = \frac{\E(S)}{\sqrt{\V(S)}},
\end{align}
where $\E(S)$ is the expectation and $\V(S)$ is the variance.
Using Eq. 16, the SNR of regular imaging is found (see Appendix A) to be 
\begin{align}
    \frac{S_R}{\sigma_{S_R}} = \frac{\lambda_s\sqrt{N}}{\sqrt{\lambda_s+\lambda_d+\frac{\lambda_d}{(P-1)}}}.
\end{align}
Using Eq. 17, the SNR of ghost imaging is found to be
\begin{align}
    \frac{S_G}{\sigma_{S_G}} &= \lambda_s\sqrt{N} \bigg[\lambda_s+3\lambda_s^2+(\lambda_s+\lambda_d)^3\nonumber\\
    &+(\lambda_s+\lambda_d)^2+2\lambda_d\lambda_s^2+2\lambda_s\lambda_d\nonumber\\
    &+\frac{(\lambda_s+\lambda_d)^3+(\lambda_s+\lambda_d)^2}{(P-1)}\bigg]^{-1/2},
\end{align}
where $\lambda_s$ is proportional to the source rate, $\lambda_d$ is proportional to the detector background rate, $P$ is the number of pixels, and $N$ is the number of realizations.

The above closed expressions for the SNR of ghost imaging and regular imaging for a single pixel hole object can be extended (not given in this work) for an object with an arbitrary number of single pixel holes by finding the expectation values and variances of Eqs. 29 and 30 in the Appendix A, and using the same mathematics to derive Eqs. 19 and 20.

\section{Results}

The results of the simulation for ghost and regular imaging are given in figures 3 and 4, along with curves generated from Eqs. 19 and 20. Figure 3 depicts the case where $\lambda_d=0$. This indicates that regular imaging is better or equal to ghost imaging. Figure 4 represents the case where the detectors have dark counts characterized by $\lambda_d=0.01$. Below $\lambda_s=0.1$, the ghost imaging SNR starts to dominate. This can be explained by Eqs. 19 and 20, where the ghost imaging technique suppresses the deleterious effects of the dark count while the regular image drowns in the dark count noise. 

\begin{figure}[ht]
   \includegraphics[width = \linewidth]{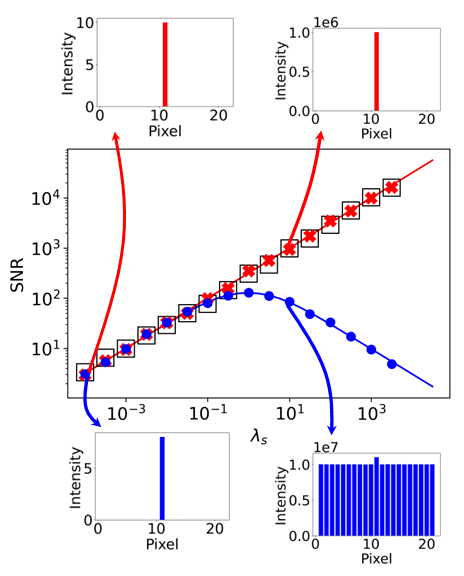} 
   \caption{Ghost Imaging and Regular Imaging without dark count. The SNR is shown as a function of the unitless number $\lambda_s=R_{s}t_{w}$ ($R_{s}$ is the rate of detected photons per pixel, and $t_{w}$ is the time window). The detector dark count is kept constant at $\lambda_d = 0$, the number of realizations $N=10^{5}$, and the number of pixels $P=21$. The data points are obtained by the simulation, where blue circular dots are for ghost imaging and red crosses are for regular imaging (black squares are gated imaging). The solid lines are obtained by the theoretical analysis (Eqs. 19 and 20). Example images are provided at certain points for comparison. In the region where $\lambda_s>0.1$, we observe that regular and gated imaging offer better SNR in comparison to ghost imaging. In the region $\lambda_s<0.1$, we observe that all three imaging methods produce the same SNR.} 
  \label{fig:resultwithoutbg}
\end{figure}

\begin{figure}[ht]
   \includegraphics[width = \linewidth]{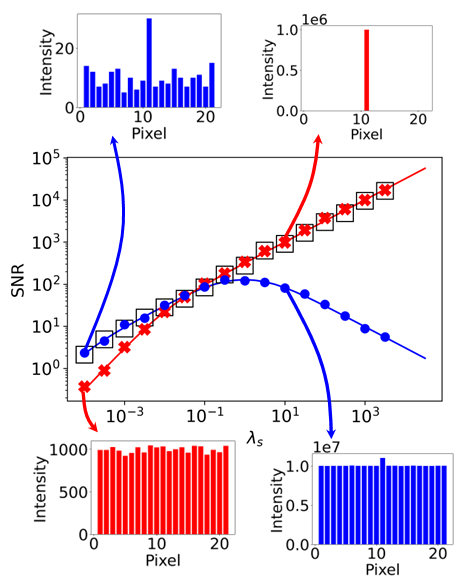} 
   \caption{Ghost Imaging and Regular Imaging with dark count. The SNR is shown as a function of $\lambda_s=R_{s}t_{w}$ ($R_{s}$ is the rate of detected photons per pixel, and $t_{w}$ is the time window). The detector dark count is kept constant at $\lambda_d = 0.01$, the number of realizations $N=10^{5}$, and the number of pixels $P=21$. The data points are obtained by the simulation, where blue circular dots are for ghost imaging and red crosses are for regular imaging (black squares are gated imaging). The solid lines are obtained by the theoretical analysis. Example images are provided at certain points for comparison. In the region where $\lambda_s>0.1$, we observe that regular and gated imaging offer better SNR in comparison to ghost imaging, as illustrated by the images of the single pixel hole object at the point $\lambda_s=10$. In the region $\lambda_s<0.1$, we observe that ghost imaging and gated imaging produce better images than regular imaging. For $\lambda_s=0.0001$, the SNR of ghost imaging is about an order magnitude better than regular imaging, as illustrated by the images at that point.} 
  \label{fig:resultwithbg}
\end{figure}

\begin{figure}[ht]
   \includegraphics[width = \linewidth]{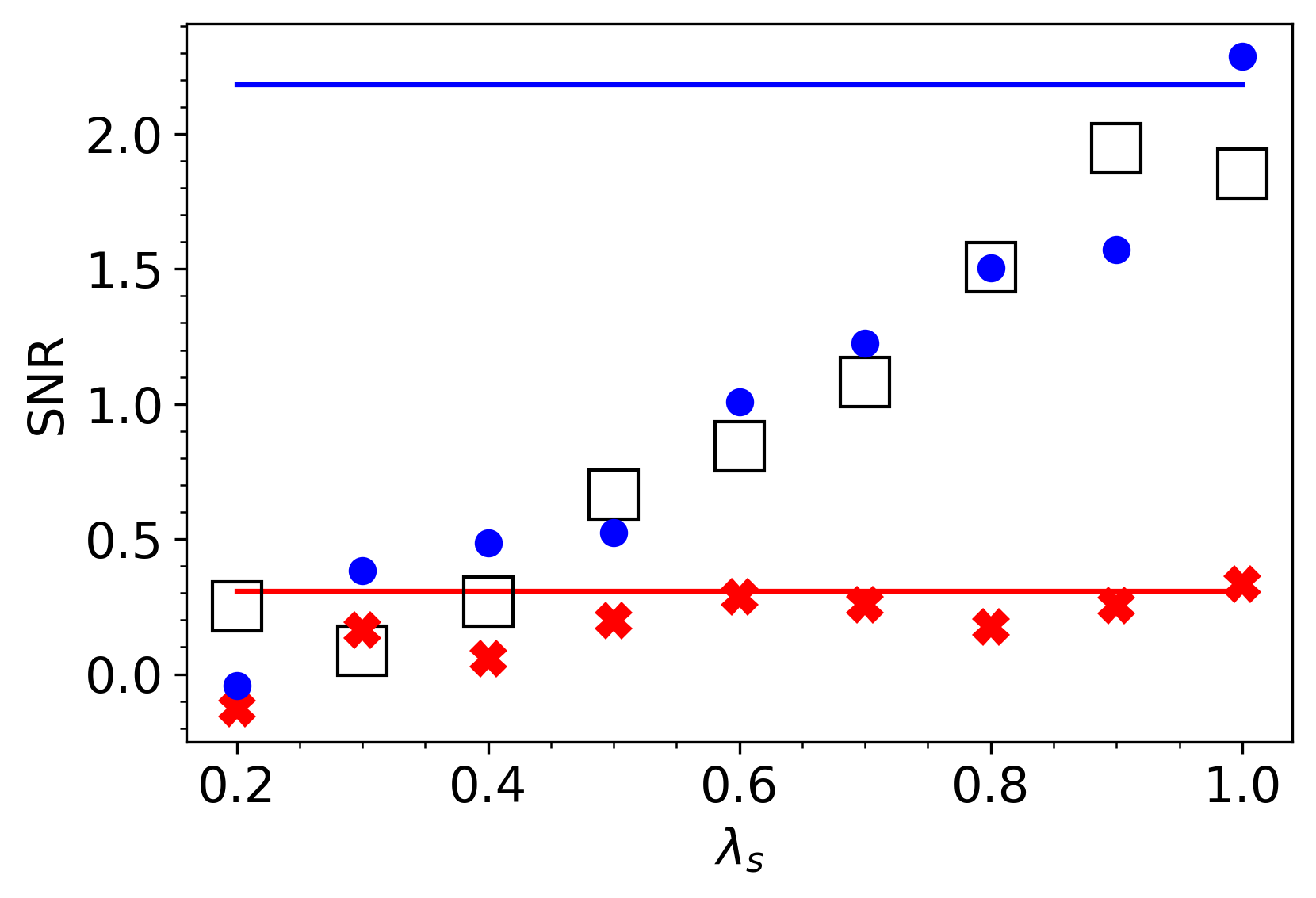} 
   \caption{Ghost Imaging and Regular Imaging as function of quantum efficiency. We chose to compare the two at the region where we observed ghost imaging to be most advantageous in figure 4 ($\lambda_s=10^{-4}$, $\lambda_d=10^{-2}$, $N = 10^5$, and $P=21$). The advantage of Ghost imaging reduces with quantum efficiency. To observe the single pixel object (with an SNR above 1) the quantum efficiency needs to exceed 0.6. The blue and red horizontal lines are the predictions for SNR given by Eqs. 19 and 20, where the quantum efficiency equals 1.} 
  \label{fig:SNRvsQE}
\end{figure}

\begin{figure}[ht]
   \includegraphics[width = \linewidth]{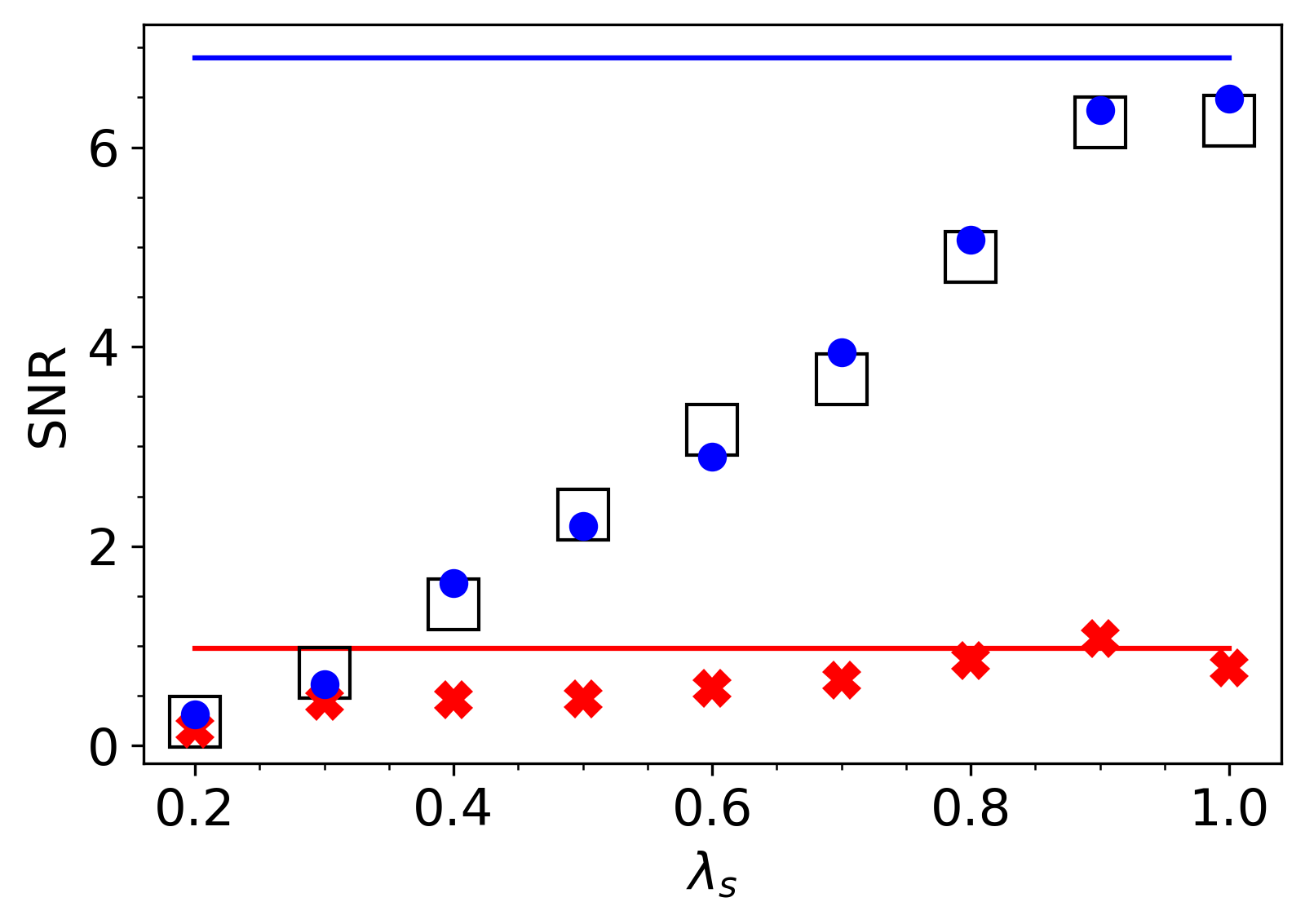} 
   \caption{Ghost Imaging and Regular Imaging as a function of quantum efficiency with increased number of realizations, $N$. All parameters are kept the same from figure 5 except $N$. Here, $N=10^6$. Ghost imaging  reveals the object for quantum efficiencies higher than 0.4. The blue and red horizontal lines are the predictions for SNR given by Eqs. 19 and 20, where the quantum efficiency equals 1.} 
  \label{fig:SNRvsQE}
\end{figure}

Another way to understand this advantage is by comparing ghost imaging to gated imaging. If we switch the roles of the bucket detector and camera in ghost imaging in figure 2, we have a gated imaging setup.\cite{morris2015imaging} In gated imaging, the camera only looks at the object (within a specified time window) when the bucket detector observes photons from the source. Hence, background noise from the detectors is suppressed in gated imaging, while that is not true for regular imaging. This background suppression also holds for ghost imaging since we are still using the time correlation between the bucket detector and camera. The black data points in figures 3 and 4 represent the SNR for gated imaging as $\lambda_s$ is varied. We can observe that the SNR of ghost imaging and gated imaging are identical at low $\lambda_s$. The regime where ghost imaging gets a significant advantage over regular imaging is $\lambda_s=10^{-4}$, $\lambda_d=10^{-2}$, $N =10^5$, and $P=21$. The regular and ghost images for these parameters are shown in figures 3 and 4. 

Given that the best real detectors for photons, such as Electron Multiplying Charge-Coupled Devices (EMCCD) and back-illuminated thinned CCDs, have a quantum efficiency of 0.9, we also studied the effect of quantum efficiency of the detectors on the SNR of our images. Figure 5 shows the SNRs of regular, ghost and gated imaging as a function of quantum efficiency, where we held $\lambda_s=10^{-4}$, $\lambda_d=10^{-2}$, $N = 10^5$, and $P=21$ constant, which are the parameters that give the best imaging advantage for ghost imaging in figure 4. As the quantum efficiency of the detectors goes down, the SNR of the ghost image also goes down. For this example, ghost imaging remains advantageous over regular imaging as long as the quantum efficiency is greater than 0.6; below 0.6, both imaging techniques cannot observe the single pixel object, because the SNR drops below one. If we increase the number of realizations $N$ to $10^6$, equivalent to multiplying the experiment run time by 10, we can observe in figure 6 that the ghost imaging advantage is clear for quantum efficiencies greater than 0.4.

A discussion of a proposed experimental design  is provided in the Supplementary Materials. Attention is given to how the SNR of the ghost image and regular image depends on the exposure of the object and the dark count rate of the detectors. 

\section{Summary, Discussion and Conclusion}
Our goal was to find a simple computational example where ghost imaging is advantageous over regular imaging. By using the SNR as a metric, we found numerically and analytically a regime that shows that advantage. We found that ghost imaging suppresses the background detector noise in a similar fashion as compared to gated imaging, while regular imaging does not.  Our closed expression for SNR depends explicitly on dark count, source intensity and number of realizations and thus can be used to shed light on experimental design requirements. We have also shown that, for moderate detector quantum efficiencies, ghost imaging remains advantageous.

The use of a single pixel hole as the object, limits the use of our study to some extent. A double-hole object simulation is provided in the supplementary materials to show that some extension of the use of SNR to other objects is possible, and also to illucidate its specific limits. Nevertheless, single pixel SNR may have use beyond these simple objects. Consider a complex image such as a skull as done in the review by Padgett and Boyd \cite{padgett2017introduction}. A study that identifies the advantages of, for example, sparse representation \cite{lane2020advantages} designed for ghost imaging of complex objects, could now switch the object to a single pixel hole and quantify the part of the image improvement that is due to the single-pixel ghost-imaging advantage. That SNR advantage can now be computationally removed to investigate the advantages of different versions of sparse representation. 

 Understanding the advantages of ghost imaging is interesting for many applications. For example, recent ghost imaging developments may find application in bio-medicine and archaeology.\cite{klein2022chemical} Here the advantage is not found in SNR and using low photon rates but in the use of compressed sensing. Quantum ghost imaging has the additional capability to detect without a bucket detector.\cite{gilaberte2019perspectives} These two examples illustrate the limitations of our exposition, as ghost imaging advantages depend on the specific task, on the object chosen, and on the experimental details of the technology used. We hope that pointing out other advantages of ghost imaging that are different in nature than the one we have discussed in some detail, helps to clarify to some extent the feature rich behavior of the phenomenon. Beyond this, the particular advantage  discussed here, that is, the improved SNR for low exposures and noisy detectors, may be of value in, for example, the context of electron ghost imaging. This recently demonstrated \cite{li2018electron} technique has the advantage that the target object could be exposed to fewer electrons. If high-resolution electron ghost imaging can be combined with new methods of protecting biological cells, such as covering the biological specimen with graphene layers,\cite{koo2020live} this may bring us closer to observing living tissue using an electron microscope.

\section{COI statement}
The authors have no conflicts to disclose.

\section{Appendix A: Mathematical Details and Outlook}
In statistics, the variance of a quantity $x$ is given by
\begin{align}
    \V(x)=\E(x^2)-(\E(x))^{2}.
\end{align}
We need Eq. 21 to solve for $\V(S)$ in Eq. 18.
Let $x_1,x_2,x_3,...,x_N$ be uncorrelated random numbers such that $\E(x_i)=\E(x_1)$ and $\E(x_i^2)=\E(x_1^2)$, where $i\in \{1,2,3,4,...,N\}$. This means $\V(x_i)=\V(x_1)$. Then, by the definition of the expectation value and Bienaymé's identity,\cite{loeve1977elementary} we get 
\begin{align}
    &\E\left(\sum_{i=1}^N x_i\right)=N\E(x_1)\\
    &\V\left(\sum_{i=1}^N x_i\right)=\sum_{i=1}^N \V(x_i)=N\V(x_1),
\end{align}
respectively.
Given that $S_{G}$ (Eq. 17) depends quadratically on $x$, the variance $\V(S)$ defined by Eq. 21 depends quartically on $x$. Therefore, the values of the raw moments for Poisson distributions up to the fourth moment are needed to evaluate Eq. 18. If $x$ is a Poisson random, dependent on $\lambda$, then
\begin{align}
    &\E(x^k)=\sum_{i=1}^k\lambda^iS_t(k,i),
\end{align}
where $S_t$ denotes Stirling numbers of the second kind.\cite{riordan1937moment} Therefore, we get
\begin{align}
    &\E(x)=\lambda\\
    &\E(x^2)=\lambda^2+\lambda\\
    &\E(x^3)=\lambda^3+3\lambda^2+\lambda\\
&\E(x^4)=\lambda^4+6\lambda^3+7\lambda^2+\lambda.
\end{align}

To determine the SNR of ghost imaging for multiple, say, $H$, single pixel holes in the object (black or white pixels, not grey pixels), we can use the expression
\begin{align}
    S_G = &\sum_{n=1}^{N}\Bigg\{\left[\sum_{j=1}^{H}X_{j,n}(\lambda_s)+Y_{b,n}(\lambda_d)\nonumber\right]\\
    &\Bigg[\sum_{j=1}^{H}\left(\frac{(X_{j,n}(\lambda_s)+Y_{j,n}(\lambda_d)}{H}\right)\nonumber\\
    &-\sum_{l=1}^{P-H}\left(\frac{(X_{l,n}(\lambda_s)+Y_{l,n}(\lambda_d)}{P-H}\right)\Bigg]\Bigg\},
\end{align}
where $Y_{b,n}$ is the bucket detector dark count, $j$ are the pixel position indices with a hole in the object, $l$ are the pixel position indices without a hole, $Y_{j,n}$ are the camera dark counts at the pixels with a hole in the object, and $Y_{l,n}$ are the camera dark counts at the pixels without a hole in the object. To determine the SNR of regular imaging for $H$ single pixel holes in the object, we can use the expression
\begin{align}
    S_R = \sum_{n=1}^{N}\Bigg[&\sum_{j=1}^{H}\left(\frac{(X_{j,n}(\lambda_s)+Y_{j,n}(\lambda_d)}{H}\right)\nonumber\\
    &-\sum_{l=1}^{P-H}\frac{Y_{l,n}(\lambda_d)}{P-H}\Bigg].
\end{align}


\section{Supplementary Materials: A double-hole object}

To investigate the range of the usefulness of the results in this paper, it is perhaps of value to consider an object that consists of two single-pixel holes, a modest extension of the single-hole object. It could also be of interest to make a pixel ``grey", in stead of black and white, by assigning a transmission value between 0 and 1 to the pixels. The question might be, if we image such a double-hole object, can we quantify by how much ghost imaging improves the quality of the image as compared with regular imaging for the same number of photons hitting the object. Such tasks could be given to students taking an optics course and are straightforward modifications of the simulation. The measure of image quality provided by the SNR is ambiguous for two not identical grey pixels, as the same value of SNR will be obtained when we switch the pixels. For more pixels this problem compounds. Nevertheless, the simulation is easily modified to calculate the ghost image and the SNR for, for example, two fully open pixels (see Fig. 7).  The simulated SNR lies above the analytic results for low $\lambda_s$. The reason is that the analytical obtained lines are those for a single hole (to facilitate comparison with figure 4). The double-hole transmits twice as many photons and one expects and finds an improvement for the SNR for low $\lambda_s$.  

\begin{figure}[ht]
   \includegraphics[width = \linewidth]{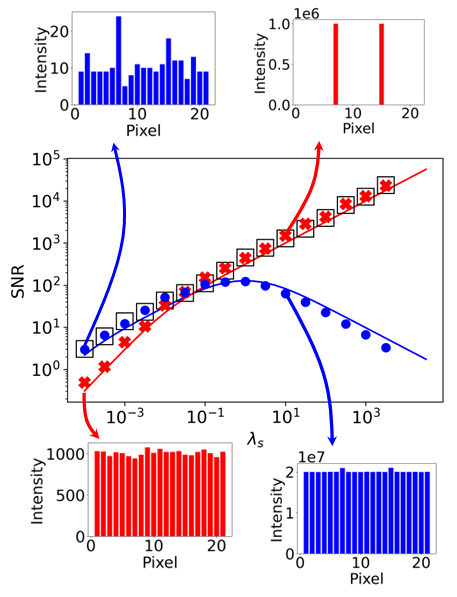} 
   \caption{Double-hole Ghost Imaging and Regular Imaging with dark count. The SNR is shown as a function of $\lambda_s$ The parameters are the same as for figure 4 in the main text. The behavior is similar to that shown in figure 4. As the analytical obtained lines are the SNR for a single hole, the simulated SNR lies above the analytic results for low $\lambda_s$. } 
  \label{fig:resultwithbg}
\end{figure}

\section{Supplementary Materials: Experimental consideration}

Given that the SPDC correlated photon source is the workhorse for quantum optics and readily available, we consider the possibility to demonstrate the above discussed ghost imaging advantage with this tool.
In SPDC, one $\hbar \omega$ photon generates two $\hbar \omega/2$ photons. The momenta of the $\hbar \omega/2$ photons transverse to the incoming photon direction are equal and opposite, resulting in momentum correlation upon the generation of the photon pair. To obtain the spatial correlation that we need for the approach we analyzed above, one can image the near-field photons.

 For a typical SPDC source, one 400 nm photon generates two 800 nm photons. Under typical operation, the avalanche photon detectors record a similar count rate, $R_{s}$, of about $10^4$/s 800 nm photons at each detector. The dark count rate of the detectors, $R_d$, is about $10^3$/s. For a measurement time window $t_w$ of the order of 10 ns, the source has $\lambda_s=R t_w$ of about $10^{-4}$. In practice, the measurement time cannot be controlled to such a short duration. Instead, one can measure for (for example) a time of $T=1$ ms and consider that to be the equivalent of $N=10^5$ realizations of 10 ns measurements for the purpose of obtaining a value for $\lambda_{s}$. The width of each of the two 800 nm beams is about 1 cm on the detector screen (placed 0.5 m away from the SPDC) and are fully separated from each other. For a movable slit of 1 mm width, placed in front of the detector, the detection rate is cut by a factor of $\epsilon = 0.1$. The number of counts $N_s$ observed per pixel for regular imaging in this example would be $N_s = R_{s} \epsilon T = (\lambda_{s} / t_w) \epsilon (t_w N)=1$. The corresponding value of $\lambda_s=R_{s}t_{w}=10^{-4}$ is a source strength that is in the regime for which the predicted imaging advantage is clear (see figure 4). The detector dark count strength is $\lambda_d=R_{d}t_{w}=10^{-5}$. In this case, a regular image would have an SNR above 1 (similar to the $\lambda_{d}=0$ case), thus the detector noise has to be increased to drown the regular image in the detector noise (as in figure 4, left lower inset). 

This can be achieved by increasing the amount of ambient light. This needs to be done in such a way that the ambient light strikes the detector directly and not the object. For $\lambda_d=10^{-2}$, the number of counts detected for regular imaging would be $N_D=R_d T=(\lambda_d/t_w) (t_w N)=1000$ (figure 4) with a rate of $R_d=10^6/s$. In this way one can match the parameters used in the above analysis ($\lambda_s=10^{-4}$, $\lambda_d=10^{-2}$, $N=10^5$).
The measured coincident rate $R_c$ is about $10^2$/s. These coincidences are measured when the two detectors are positioned symmetrically with respect to the axis defined by the photon incident on the SPDC. The measured coincident rate $R_c$ could be as high as $\epsilon R_{s}=10^3$/s if both detectors would, without fail, record every photon pair. If one of the detectors is moved laterally by 5 mm, the observed coincidence rate drops to $1$/s, while the singles rate stays nominally the same. 

\begin{figure}[ht]
   \includegraphics[width = \linewidth]{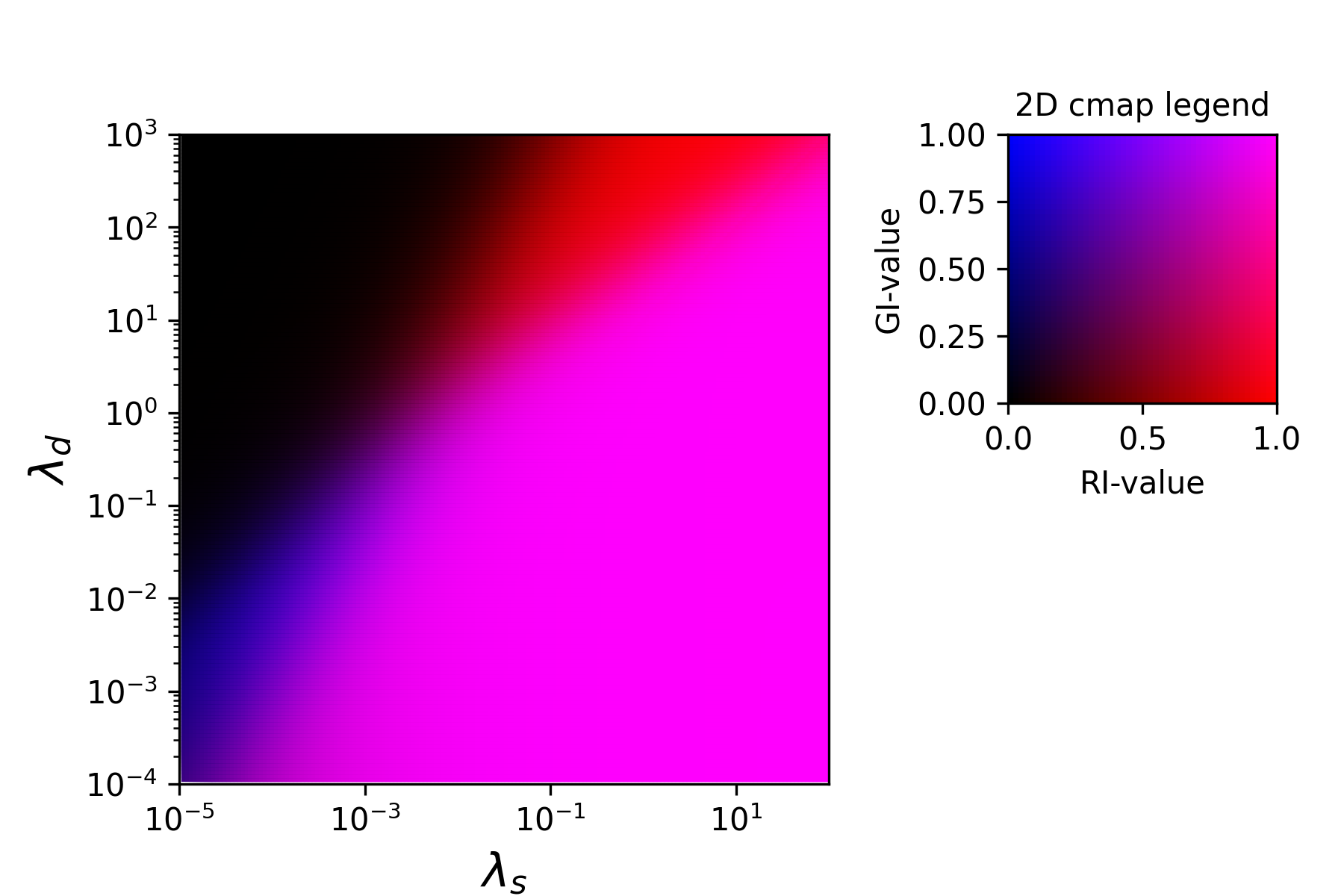} 
   \caption{SNR parameter space. For experimental consideration it may be of use to consider the SNR as a function of $\lambda_s$ and $\lambda_d$ ($N=10^{5}$). Here, the SNRs are graphically compared by a GI-value and RI-value, which  range from 0 to 1. The GI-value $=\text{SNR}_{G}/(1+\text{SNR}_{G})$ and RI-value $=\text{SNR}_{R}/(1+\text{SNR}_{R})$, where $\text{SNR}_{G}$ and $\text{SNR}_{R}$ are the SNRs for ghost images and regular images, respectively, for a single pixel hole object. A value approaching 1 means that the SNR is much larger than 1 and the image is clear. A value of 0 means the SNR is much smaller than one and the image is not visible. The GI- and RI-values are plotted against $\lambda_s$ and $\lambda_d$ (both axes are log scale), while keeping $N=10^5$ realizations and $P=21$ pixels constant. Blue means one gets only a clear ghost image; red means only a clear regular image; purple means both images are good; black indicates poor image quality for both techniques.} 
  \label{fig:resultwithoutbg}
\end{figure}

The measured coincident background rate $N^{(2)}_{bg}$ due to randomly arriving photons (without detection slits) is given by $N^{(2)}_{bg} = R^2 t_w = 1/s$. When the detectors noise is raised with ambient light  so that again $\lambda_d=10^{-2}$, this number increases to $N^{(2)}_{bg} = 10^4/s$. This means that the SNR for regular imaging and ghost imaging is the same. This rather constructed, but nevertheless representative example indicates that a judicious choice of the slit width, careful alignment of the detectors, temporal and spatial compensation plates \cite{altepeter2005phase} could be used to maximize the measured coincidence rate and that an experimental demonstration of this SNR ghost imaging advantage is not trivial, but perhaps within reach. To design an experiment, it may be of use to vary the parameter $\lambda_d$ in addition to $\lambda_s$. Figure 8 provides a false color image to compare the SNR of ghost imaging to regular imaging based on Eq. 19 and 20. This indicates for which values in the parameter space each is dominant and where the SNR values exceed one, so that a clear image can be obtained.  

\section{Acknowledgements}   This work has been funded by NSF Grant No. NSF-
2207697.


%

\end{document}